\documentclass[
    aps,
    prd,
    notitlepage,
    nofootinbib,
    10pt,
    a4paper,
    preprintnumbers,
    superscriptaddress
]{revtex4-2}
\usepackage[utf8]{inputenc}
\usepackage[english]{babel}
\usepackage{microtype}
\usepackage{amsmath}
\usepackage{graphicx}

\usepackage[usenames,svgnames]{xcolor}
\usepackage{booktabs, multirow}
\usepackage[allcolors=black]{hyperref}
\usepackage{cleveref,csquotes,acronym}
\usepackage{float}
\usepackage{ulem}
\usepackage{placeins}[subsection]

\usepackage{tikz}
\usepackage{pgfplots}

\usetikzlibrary{
    external,
    fadings
}
\tikzfading[name=fade fill left,left color=transparent!100, right color=transparent!75]
\tikzfading[name=fade line left,left color=transparent!100, right color=transparent!25]
\usepgfplotslibrary{
    groupplots,
    external
}
\pgfplotsset{
    scale only axis,
    width=0.5\textwidth,
    height=0.2\textheight,
    compat = newest,
    every axis plot/.append style={
        line join=round,
        line cap=round,
        clip=false,
    },
    x axis line style = {thick},
    y axis line style = {thick},
    legend style={draw=none,fill=none,},
    legend style={/tikz/every even column/.append style={column sep=0.3cm}},
}
\pgfplotsset{
    colormap={reversePuOr}{
    rgb255=(45, 0, 75)
    rgb255=(84, 39, 136)
    rgb255=(128, 115, 172)
    rgb255=(178, 171, 210)
    rgb255=(216, 218, 235)
    rgb255=(247, 247, 247)
    rgb255=(254, 224, 182)
    rgb255=(253, 184, 99)
    rgb255=(224, 130, 20)
    rgb255=(179, 88, 6)
    rgb255=(127, 59, 8)
    },
    colormap name=reversePuOr,
}

\definecolor{ggTableauTenOne}{HTML}{4e79a7}
\definecolor{ggTableauTenTwo}{HTML}{f28e2b}
\definecolor{ggTableauTenThree}{HTML}{e15759}
\definecolor{ggTableauTenFour}{HTML}{76b7b2}
\definecolor{ggTableauTenFive}{HTML}{59a14f}
\definecolor{ggTableauTenSix}{HTML}{edc948}
\definecolor{ggTableauTenSeven}{HTML}{b07aa1}
\definecolor{ggTableauTenEight}{HTML}{ff9da7}

\acrodef{ah}[AH]{apparent horizon}
\acrodef{gr}[GR]{general relativity}
\acrodef{mmg}[MMG]{minimally modified gravity}
\acrodef{mots}[MOTS]{marginally outer trapped surface}
\acrodef{gws}[GWs]{gravitational waves}
\acrodef{dof}[DoF]{degrees of freedom}

\crefname{subsection}{subsection}{subsections}
\crefname{figure}{figure}{figures}

\begin{document}

\preprint{YITP-23-75, IPMU23-0023}
\title{Spherical scalar collapse in a type-II minimally modified gravity}

\author{Atabak Fathe Jalali}
\email{fathe.atabak@yukawa.kyoto-u.ac.jp}
\affiliation{KTH Royal Institute of Technology, SE-100 44 Stockholm, Sweden}
\affiliation{Center for Gravitational Physics and Quantum Information (CGPQI),
Yukawa Institute for Theoretical Physics (YITP),
Kyoto University, 606-8502, Kyoto, Japan}

\author{Paul Martens}
\email{paul.martens@yukawa.kyoto-u.ac.jp}
\affiliation{Center for Gravitational Physics and Quantum Information (CGPQI),
Yukawa Institute for Theoretical Physics (YITP),
Kyoto University, 606-8502, Kyoto, Japan}

\author{Shinji Mukohyama}
\email{shinji.mukohyama@yukawa.kyoto-u.ac.jp}
\affiliation{Center for Gravitational Physics and Quantum Information (CGPQI),
Yukawa Institute for Theoretical Physics (YITP),
Kyoto University, 606-8502, Kyoto, Japan}
\affiliation{Kavli Institute for the Physics and Mathematics of the Universe (WPI),
The University of Tokyo, Kashiwa, Chiba 277-8583, Japan}

\date{\today}

\begin{abstract}
We investigate the spherically-symmetric gravitational collapse of a massless scalar field in the framework of a type-II minimally modified gravity theory called VCDM.
This theory propagates only two local physical \ac{dof} supplemented by the so-called instantaneous (or shadowy) mode.
Imposing asymptotically flat spacetime in the standard Minkowski time slicing, one can integrate out the instantaneous mode. Consequently, the equations of motion reduce to those in general relativity (GR) with the maximal slicing. Unlike GR, however, VCDM lacks 4D diffeomorphism invariance, and thus one cannot change the time slicing that is preferred by the theory. We then numerically evolve the system to see if and how a black hole forms. For small amplitudes of the initial scalar profile, we find that its collapse does not generate any black hole, singularity or breakdown of the time slicing. For sufficiently large amplitudes, however, the collapse does indeed result in the formation of an apparent horizon in a finite time. After that, the solution outside the horizon is described by a static configuration, \textit{i.e.} the Schwarzschild geometry with a finite and time-independent lapse function. Inside the horizon, on the other hand, the numerical results indicate that the lapse function keeps decreasing towards zero so that the central singularity is never reached. This implies the necessity for a UV completion of the theory to describe physics inside the horizon. Still, we can conclude that VCDM is able to fully describe the entire time evolution of the Universe outside the black hole horizon without knowledge about such a UV completion.
\end{abstract}

\maketitle

\section{Introduction}

When \ac{gr} was first formulated, it showed promise by solving observational riddles that deprived physicists of their sleep at the time. Even today, \ac{gr} continues to produce physically accurate predictions, \textit{e.g.} the detection of \ac{gws} by LIGO/Virgo \cite{LIGOScientific:2016aoc, LIGOScientific:2021qlt} and the first-ever photograph of a black hole shadow \cite{EventHorizonTelescope:2019dse}. However, albeit a simple yet rigorous, elegant, and symmetrically satisfying theory, \ac{gr} has proven incomplete on quantum gravity scales, and some observations may currently be bearing witness to its failure on cosmological scales. Once more, physicists find themselves in a state of sleep deprivation.

This motivates us to look beyond \ac{gr}. Do we need a completely new theory? Is it possible to modify gravity in a way such that it includes a complete description of quantum gravity, and breeds a cosmological framework that is consistent with observations? The experiments conducted at LIGO/Virgo open up the possibility of probing strongly gravitating events through \ac{gws}, such as black hole binaries and gravitational collapse. This allows for further testing of the validity of \ac{gr} and alternative frameworks.

Several such alternative theories of gravity have already been proposed. In the high-energy limit, superstring theories \cite{Polchinski:1998rr} and Ho\v{r}ava-Lifshitz gravity \cite{Horava:2009uw} are examples of candidates for a theory of quantum gravity, while massive gravity, bimetric gravity, and various scalar-tensor theories \cite{deRham:2010kj,Schmidt-May:2015vnx,PhysRevD.37.3406,Arkani-Hamed:2003pdi,Horndeski:1974wa} carry implications for the dark sector of the Universe as well as its accelerated expansion. A common artifact when modifying gravity is accompanying \ac{dof}. On cosmological scales, these are exploited to explain observational discrepancies. However, on astrophysical scales, these must also avoid contradicting well-established experimental data and the appearance of various instabilities such as ghosts \cite{DeFelice:2006pg}. One common method of dealing with this is to implement different screening mechanisms \cite{terHaar:2020xxb}. Alternatively, one could consider modifying gravity \textit{minimally}, \textit{i.e.} keeping only two local physical \ac{dof}. Such a theory is denoted a \ac{mmg} theory~\cite{Lin:2017oow}.
This class of theories defines itself by not introducing any additional local physical degrees of freedom other than those in \ac{gr}. Therefore, they easily avoid the aforementioned instabilities and constraints that could come from the extra propagating degrees of freedom that are commonly found in other modified gravity theories. And this is the case even without needing any screening mechanisms.

It was previously established in \cite{Aoki:2018brq} that all \ac{mmg} theories may be divided into two types: type-I and type-II. The former are theories that are equivalent to \ac{gr} in the absence of matter but that modify gravity due to non-trivial matter coupling \cite{Aoki:2018zcv}, while the latter are theories simply different from \ac{gr}, even in the absence of matter. For example, the class of \ac{mmg}s studied in \cite{Lin:2017oow} were, by \cite{Carballo-Rubio:2018czn}, mostly shown to be of type-I, and the most generic construction of a subclass of type-I \ac{mmg} with the standard dispersion relation for the tensorial \ac{gws} is elaborated upon in subsection IV A of \cite{Aoki:2021zuy}. Another example would be \cite{Feng:2019dwu}. Instances of type-II \ac{mmg} theories include the Cuscuton \cite{Afshordi:2006ad}, the minimal theory of massive gravity \cite{DeFelice:2015hla}, a consistent $D\to 4$ Einstein-Gauss-Bonnet gravity \cite{Aoki:2020lig}, and VCDM \cite{DeFelice:2020eju} (see also \cite{Yao:2020tur,Yao:2023qjd}).

We have chosen to further investigate VCDM \cite{DeFelice:2020eju}. Unlike \ac{gr}, VCDM does not have full 4D diffeomorphism invariance (although 3D spatial diffeomorphism invariance still holds) and effectively substitutes the cosmological constant with a free potential function $V(\phi)$ of a non-propagating auxiliary scalar field $\phi$ in the gravitational action.
This last feature gives \enquote{VCDM} its name. The theory is kept minimal thanks to the built-in constraints that are imposed on the field $\phi$. Consequently, VCDM is ghost-free, as the only propagating modes are the standard tensor modes, without requiring any screening mechanisms are needed. Nonetheless, the theory does admit a phenomenology that is different from \ac{gr}. Indeed, so far, VCDM has proven to be a fruitful framework within which to approach multiple questions, such as tensions in late-time cosmology \cite{DeFelice:2020eju,DeFelice:2020cpt,DeFelice:2020prd} and a bouncing Universe \cite{Ganz:2022zgs}. Stars \cite{DeFelice:2021xps}, as well as black holes and gravitational collapse \cite{DeFelice:2020onz,DeFelice:2021xps} have also been studied. A thorough comparison with the Cuscuton has been performed in \cite{ Aoki:2021zuy,DeFelice:2022uxv}, the latter showing how any solution of Cuscuton is a solution of VCDM, while the converse is not necessarily true. It was further shown that there exists a subset of solutions in VCDM that coincide with exact solutions in \ac{gr}, given specific conditions on the foliation of spacetime.

In this paper, we are interested in numerically simulating the gravitational collapse of a massless scalar field in VCDM. It is well-established in \ac{gr} that black holes form after a gravitational collapse. By virtue of its aforementioned overlap with \ac{gr} solutions, one may expect VCDM to admit black holes from gravitational collapse as well, at least in specific foliations of spacetime. On the other hand, unlike \ac{gr}, the fundamental symmetry of VCDM is not the 4D diffeomorphism invariance but the invariance under the foliation-preserving diffeomorphism, and thus one cannot change the time slicing that is preferred by the theory. In VCDM, different time foliations represent physically different solutions, and whether or not an \ac{ah} appears before a singularity and a breakdown of the time foliation carries great importance. A collapsing cloud of dust was recently subject to a numerical study and did indeed result in the formation of a black hole \cite{DeFelice:2022riv}, its solution coinciding with a particular foliation of the Oppenheimer-Snyder collapse in which the \ac{ah} forms prior to the singularity and the breakdown of the foliation.

The structure of this paper is as follows. \Cref{sec:an_overview_of_vcdm} gives a brief overview of VCDM as a theory of gravity and formulate the total action in the unitary gauge. In \cref{sec:setup}, assuming a spherically symmetric ansatz, we derive the equations of motion. These are simplified by demanding an asymptotically flat spacetime, allowing us to integrate out the instantaneous (or shadowy) mode and acquire a constant $\phi$ and traceless extrinsic curvature. After specifying relevant boundary conditions and initial conditions, the equations of motion are ready to be integrated. The numerical setup, as well as the simulation results, are presented in \cref{sec:numerical_integration}. We there confirm the correct behavior of the numerical convergence, the appearance of an \ac{ah}, and the evolution of the lapse function inside the \ac{ah} towards zero. Finally, \cref{sec:summary_and_discussion} provides a summary of the paper and a discussion of the obtained results, concluding with some suggestions for future projects.

\section{An overview of VCDM}\label{sec:an_overview_of_vcdm}

We consider a type-II \ac{mmg} theory called VCDM, which was first formulated in \cite{DeFelice:2020eju}\footnote{In particular, section 2 for the complete construction of VCDM, and section 5 for its phenomenology that differs from \ac{gr}}. The construction of VCDM begins with a canonical transformation of vacuum \ac{gr} in the ADM formalism. The generating functional $F$ for this canonical transformation is expressed in terms of two quantities $\phi$ and $\psi$. These are equal to some combinations of the transformed and untransformed canonical coordinates. They are later promoted to non-dynamical fields and constrained to evaluate to their respective expressions of the canonical coordinates. A cosmological constant term is then introduced in the frame after the transformation. This demotes the Hamiltonian constraint from first-class to second-class and the theory is now different from \ac{gr}. However, as a consequence of the demotion, the number of the phase space \ac{dof} increases to $5$, and the addition of a gauge-fixing term to the action is necessary so as to reduce the dimension of the constraint surface in phase space down to $4$ again. The broken symmetry that was originally generated by the Hamiltonian constraint is time diffeomorphism invariance; in exchange, the theory is different from \ac{gr}, yet propagates only two physical \ac{dof}.
The theory is also ghost-free, simply
because there is no extra gravitational mode besides the standard tensor modes.
Thus, it avoids having to employ screening mechanisms, and yet we still have new phenomenology on cosmological scales.

In the ADM formalism, spacetime is foliated by a family of $t$-constant non-intersecting 3D space-like hypersurfaces. Choosing a set of coordinates $(t,x^i)$ ($i=1,2,3$), the line element takes the form
\begin{equation}\label{eq:ADM_line-element}
    g_{\mu\nu}dx^{\mu}dx^{\nu} = -N^2dt^2+\gamma_{ij}(N^idt+dx^i)(N^jdt+dx^j),
\end{equation}
where $N$ is the lapse function, $N^i$ is the shift vector, and $\gamma_{ij}$ (with inverse $\gamma^{ij}$) is the induced 3D spatial metric on the constant $t$ hypersurface. In contrast to \ac{gr} and as a consequence of the theory's construction, the Lagrangian density of VCDM additionally depends on an auxiliary scalar field $\phi$, a free function $V(\phi)$, and two Lagrange multipliers $\lambda$ and $\lambda_{\rm gf}^i$. By working in the so-called unitary gauge, the gravitational action of VCDM can explicitly be written
\begin{equation}\label{eq:grav_action_vcdm}
    I_{\rm g}=M_{\rm Pl}^{2}\int dt d^3x\,N\sqrt{\gamma}\left\{\frac{1}{2}\,\left[R+K_{ij}K^{ij}-K^{2}-2V(\phi)\right]-\frac{\lambda_{{\rm gf}}^{i}}{N}\,\partial_{i}\phi-\frac{3}{4}\,\lambda^{2}-\lambda\,(K+\phi)\right\}\,,
\end{equation}
where $M_{\text{Pl}}=1/\sqrt{8\pi G_N}$ is the reduced Planck mass, $G_N$ denotes Newton's gravitational constant, $\gamma$ is the determinant of $\gamma_{ij}$, and $R$ denotes the three-dimensional Ricci scalar of $\gamma_{ij}$. The extrinsic curvature $K_{ij}$ of the $t$-constant hypersurfaces, its inverse $K^{ij}$, and its trace $K$ are defined by
\begin{align}
    K_{ij} &= \dfrac{1}{2N}(\partial_t\gamma_{ij}-D_{i}N_{j}-D_{j}N_{i}),\\
    K^{ij}&=\gamma^{ik}\gamma^{jl}K_{kl},\\
    K&=\gamma^{ij}K_{ij}.
\end{align}
Here, $D_{i}$ is the three-dimensional covariant derivative compatible with $\gamma_{ij}$, and $N_i=\gamma_{ij}N^j$. Finally, we mention that the two Lagrange multipliers $\lambda$ and $\lambda^i_{\rm gf}$ in equation \eqref{eq:grav_action_vcdm} serve to constrain the field $\phi$ so that the theory remains minimal, \textit{i.e.} having only two local physical \ac{dof}.

As a matter source, we consider a canonical massless scalar field $\psi$ described by the action
\begin{equation}\label{eq:matter_action}
 I_{\rm m} = \frac{M_{\rm Pl}^2}{2} \int dt d^3x N\sqrt{\gamma} \left[ (\partial_{\perp}\psi)^2 - \gamma^{ij}\partial_i\psi\partial_j\psi\right]\,,\quad \partial_{\perp} \equiv \frac{1}{N}(\partial_t - N^i\partial_i)\,.
\end{equation}
Combining equations \eqref{eq:grav_action_vcdm} and \eqref{eq:matter_action} produces the total action
\begin{equation}\label{eq:total_action}
 I_{\rm tot} = I_{\rm g} + I_{\rm m}\,.
\end{equation}

\section{Setup}\label{sec:setup}

This section is dedicated to deriving the equations of motion and preparing these for the numerical integration (\cref{sec:numerical_integration}). We begin in \cref{sec:basic_equations} by assuming a spherically symmetric ansatz and then deriving the equations of motion. In \cref{sec:integrating_out_shadowy_mode}, we impose asymptotic flatness in the standard Minkowski slicing and integrate out the so-called instantaneous (or shadowy) mode. Consequently, all solutions of this system will be \ac{gr} solutions in the maximal slicing (\textit{i.e.} $K=0$). \Cref{sec:Nondynamical_Q} downgrades $Q$ to a non-dynamical variable, which is practical for numerical stability. We finally conclude by specifying the boundary conditions and initial conditions for all variables in \cref{sec:Boundary_conditions,sec:Initial_condition}, respectively.

\subsection{Basic equations}\label{sec:basic_equations}

We adopt the spherically symmetric ansatz
\begin{equation}
 N=\alpha(t,r)\,, \quad N_idx^i = \beta(t,r)dr\,, \quad \gamma_{ij}dx^idx^j = dr^2 + \Phi(t,r)^2d\Omega_2^2\,, \quad \psi = \psi(t,r)\,.
\end{equation}
where $\alpha$ and $\beta$ now relate to the lapse and the shift, respectively, and $\Phi$ is the areal radius.
In parallel, we also write
\begin{equation}
 \phi=\phi(t,r)\,, \quad \lambda=\lambda(t,r)\,, \quad \lambda_{{\rm gf}}^i\partial_i = \tilde{\lambda}(t,r)\partial_r\,,
\end{equation}
where $d\Omega_2^2$ is the metric of the unit $2$-sphere. In order to simplify the equations of motion, we will use the trace of the extrinsic curvature that now reads
\begin{equation}
 K = 2\partial_{\perp}\ln\Phi - \frac{1}{\alpha}\partial_r\beta\,,
\end{equation}
where $\partial_{\perp} = (1/\alpha)(\partial_t-\beta\partial_r)$. Writing $K^i_j\equiv \gamma^{ik}K_{kj}$, we also introduce the following variables
\begin{equation}
 Q \equiv K^r_r - \frac{1}{3}K = -\frac{2}{3}\left(\partial_{\perp}\ln\Phi + \frac{1}{\alpha}\partial_r\beta\right)\,, \quad P \equiv \partial_{\perp}\psi\,, \quad a \equiv \partial_r\ln\alpha\,.
 \label{eq:Definition_of_Q_P_and_a}
\end{equation}

Given this ansatz, the position of an \ac{ah} $r=r_{\rm AH}$, can be found by solving\footnote{A derivation of this condition can be found in \cref{appendix:A}. For more details, please consult subsection III B in \cite{DeFelice:2022riv} and/or subchapters 2.4 and 5.1.7 in \cite{Poisson:2009pwt}.}
\begin{equation}
    \label{eq:AH_condition}
 \left. g^{\Phi\Phi}\right|_{r=r_{\rm AH}} = 0\,,
\end{equation}
where
\begin{equation}
    \label{eq:Definition_of_gPhiPhi}
 g^{\Phi\Phi} \equiv -(\partial_{\perp}\Phi)^2 + (\partial_r\Phi)^2 = -\left(\frac{1}{2}Q + \frac{1}{3}K\right)^2\Phi^2 + (\partial_r\Phi)^2\,.
\end{equation}
In terms of the introduced variables, the equations of motion can then be divided into two sets: one set with the dependence on the Lagrange multipliers $\lambda$ and $\tilde{\lambda}$ removed, and the other set determining $\lambda$ and $\tilde{\lambda}$.

The first set of equations of motion, \textit{i.e.} those independent of the Lagrange multipliers $\lambda$ and $\tilde{\lambda}$, consists of constraints, dynamical equations, and non-dynamical equations.
The constraints are
\begin{align}
 \frac{\partial_r^2\Phi}{\Phi} &=\frac{1-(\partial_r\Phi)^2}{2\Phi^2} - \frac{3}{8}Q^2 - \frac{1}{4}P^2 - \frac{1}{4}(\partial_r\psi)^2 - \frac{1}{2}V(\phi) + \frac{1}{6}\phi^2\,, \label{eq:constraint-Phi}\\
 \partial_r Q &= -3Q\partial_r\ln\Phi + P\partial_r\psi\,,\label{eq:constraint-Q}\\
 \partial_r\phi &= 0\,, \label{eq:constraint-phi}
\end{align}
where equation \eqref{eq:constraint-Phi} comes from the Hamiltonian constraint, equation \eqref{eq:constraint-Q} from the momentum constraint, and equation \eqref{eq:constraint-phi} follows from the variation with respect to $\tilde{\lambda}$. The dynamical equations are
\begin{align}
 \partial_t \psi &= \alpha P + \beta\partial_r\psi\,, \label{eq:dpsidt}\\
 \partial_t P &= \alpha \left[- KP + \partial_r^2\psi + (a +2\partial_r\ln\Phi)\partial_r\psi\right] + \beta\partial_r P\,, \label{eq:dPdt}\\
 \partial_t \Phi &= \alpha\left(\frac{1}{3}K-\frac{1}{2}Q\right)\Phi + \beta\partial_r\Phi\,, \label{eq:dPhidt}\\
 \begin{split}
     \partial_t Q &= \alpha\left[ -KQ - \frac{1}{4}Q^2 + \frac{2}{3}(a^2+\partial_r a-a\partial_r\ln\Phi) +\frac{1-(\partial_r\Phi)^2}{\Phi^2} - \frac{1}{6}P^2 + \frac{1}{2}(\partial_r\psi)^2\right.\\
    &\quad\qquad \left. + \frac{1}{9}\phi^2-\frac{1}{3}V(\phi)\right] + \beta (P\partial_r\psi-3Q\partial_r\ln\Phi),\label{eq:dQdt}
 \end{split}\\
 \partial_t \phi &= \alpha\left[ - a^2 - \partial_r a - 2a\partial_r\ln\Phi  +\frac{3}{2}Q^2 + P^2 +\frac{1}{3}\phi^2 - V(\phi)\right]\,.  \label{eq:dphidt}
\end{align}
Two non-dynamical equations are derived from the definitions of $(a, Q, K)$ and give the first spatial derivative of the lapse and the shift as
\begin{align}
 \partial_r \ln\alpha &= a\,, \label{eq:lapse} \\
 \partial_r \beta &= -\alpha \left(Q+\frac{1}{3}K\right)\,. \label{eq:shift}
\end{align}
The remaining non-dynamical equations are obtained by requiring that the time derivatives of the constraints be consistent with the spatial derivatives of the dynamical equations as
\begin{align}
 \partial_r^2 K &= -2(a+\partial_r\ln\Phi)\partial_r K - \left[ a^2 + \partial_r a + 2a\partial_r\ln\Phi - \frac{3}{2}Q^2 - P^2 - \frac{1}{3}\phi^2 + V(\phi) \right] \left[K + \phi - \frac{3}{2}V'(\phi) \right]\,, \label{eq:shadowy-K}\\
 \begin{split}
     \partial_r^2 a &= - 3a\partial_r a + a \left[\frac{(\partial_r\Phi)^2-1}{\Phi^2} + \frac{9}{4}Q^2 + \frac{3}{2}P^2+\frac{1}{2}(\partial_r\psi)^2\right] - 2a^2\partial_r\ln\Phi - a^3\\
&\qquad+ P(3Q\partial_r\psi + 2\partial_r P) - 9Q^2\partial_r\ln\Phi + 2(a\partial_r\ln\Phi - \partial_r a )\partial_r\ln\Phi\,. \label{eq:shadowy-a}
 \end{split}
\end{align}
Finally, the extra equations that determine the Lagrange multipliers $\lambda$ and $\tilde{\lambda}$ are
\begin{equation}\label{eq:EoM_lambdas}
 \lambda = -\frac{2}{3}(K+\phi)\,,\quad \partial_r\tilde{\lambda} + 2\tilde{\lambda}\partial_r\ln\Phi = -\left[\frac{2}{3}(K+\phi)-V'(\phi) \right]\alpha\,.
\end{equation}

We are interested in evolving the nine variables $(\psi, P, \Phi, Q, \phi, \alpha, \beta, K, a)$, and note that $\lambda$ and $\tilde{\lambda}$ do not appear in any of the equations \eqref{eq:constraint-Phi}-\eqref{eq:shadowy-a}. Hence, determining them will not be necessary for the later numerical integration; we shall not consider them anymore.

\subsection{Integrating out the shadowy mode}\label{sec:integrating_out_shadowy_mode}

In actual numerical studies, we restrict our considerations to the situation where the solution far from the center approaches a flat spacetime in the standard Minkowski slicing without excitations of the scalar field $\psi$. In vacuum (or for $\psi=\text{const.}$), the theory admits a flat spacetime in the standard Minkowski slicing without spontaneous breaking of the spatial diffeomorphism invariance (and thus with $\tilde{\lambda}=0$) if and only if $\phi=\phi_0$, where $\phi_0$ is a constant that simultaneously satisfies
\begin{align}
 V'(\phi_0) &=\frac{2}{3}\phi_0\,, \label{eq:phi0}\\
 V(\phi_0) &=\frac{1}{3}\phi_0^2\,. \label{eq:Lambdaeff=0}
\end{align}
One can consider equation \eqref{eq:phi0} as the definition of $\phi_0$ and equation \eqref{eq:Lambdaeff=0} as the condition setting the effective cosmological constant to zero. The latter can be achieved by tuning a constant in $V(\phi)$. Since equation \eqref{eq:constraint-phi} says that $\phi$ is independent of $r$, the assumed asymptotic condition that the solution should approach the flat spacetime far from the center implies that $\phi=\phi_0$ everywhere, where $\phi_0$ is defined by equation \eqref{eq:phi0}, and that the effective cosmological constant should be set to zero as in equation \eqref{eq:Lambdaeff=0}.

Upon introducing this change into equations \eqref{eq:constraint-Phi}-\eqref{eq:shadowy-a}, the equations of motion simplify. We note that the previously dynamical equation \eqref{eq:dphidt} for $\phi$ downgrades to a non-dynamical one for $a$, explicitly
\begin{equation}
    \partial_r a = - a^2  - 2a\partial_r\ln\Phi  +\frac{3}{2}Q^2 + P^2 \,.  \label{eq:eq_a_phi0}
\end{equation}
Because of this, we no longer include equation \eqref{eq:shadowy-a} in the set of equations to be solved; it automatically follows from other equations.

Furthermore, the equation \eqref{eq:shadowy-K} for $K$ simplifies to
\begin{equation}
    \partial_r^2 K = -2(a+\partial_r\ln\Phi)\partial_r K\,, \label{eq:shadowy-K_phi0}
\end{equation}
which can be integrated once to give $\partial_r K(t,r) = f_1(t)/[\alpha(t,r) \Phi(t,r)]$, where $f_1(t)$ is an arbitrary function of $t$. The regularity of the solution at the center of spherical symmetry, where $\Phi(t,r)=0$, requires that $\alpha(t,r=0)\ne 0$ and that $\partial_r K(t,r=0)=0$. Therefore, the regularity of the center sets $f_1(t)=0$ and thus $\partial_r K(t,r)=0$ and $K(t,r)=f_0(t)$, where $f_0(t)$ is an arbitrary function of $t$. As stated previously, we assume that the solution far from the center approaches a flat spacetime in the standard Minkowski slicing without excitations of the scalar field $\psi$. This in particular implies that $K(t,r)$ should vanish at the spatial infinity, meaning that $f_0(t)=0$ and that $K(t,r)=0$. In this way we have successfully integrated out the so-called shadowy mode, avoiding the corresponding boundary value problem.

We also point out that, as $K$ has now reduced to a constant (actually zero), equation \eqref{eq:Definition_of_gPhiPhi} can be re-written
\begin{equation}
    g^{\Phi\Phi} \equiv -(\partial_{\perp}\Phi)^2 + (\partial_r\Phi)^2 = -\frac{1}{4} Q^2\Phi^2 + (\partial_r\Phi)^2\,.
    \label{eq:gPhiPhi_without_K}
\end{equation}
In addition, $K$ vanishing implies from equation \eqref{eq:EoM_lambdas} that $\lambda$ also becomes $0$. Accordingly, the solutions to the equations \eqref{eq:constraint-Phi}-\eqref{eq:shadowy-a} after this simplification will, in fact, be exact \ac{gr} solutions; we are interested in whether or not an \ac{ah} appears, and if it does, before or after a breakdown of the time foliation and formation of a singularity. We reiterate that in \ac{gr}, a change of the spacetime foliation could reverse the order of these events. However, different foliations in VCDM correspond to physically different solutions. Any singularity or breakdown of the time foliation appearing in this theory cannot be circumvented by a change of the time slicing.

\subsection{Nondynamical $Q$}
\label{sec:Nondynamical_Q}

For actual numerical simulation, one can downgrade $Q$ from a dynamical variable to a non-dynamical one, abandoning equation \eqref{eq:dQdt} and instead using equation \eqref{eq:constraint-Q} to determine $Q$. The simplifications in the previous subsection along with such a downgrade give the equations of motion in their final form. Only the following constraint equation remains
\begin{equation}
 \frac{\partial_r^2\Phi}{\Phi} =\frac{1-(\partial_r\Phi)^2}{2\Phi^2} - \frac{3}{8}Q^2 - \frac{1}{4}P^2 - \frac{1}{4}(\partial_r\psi)^2\,, \label{eq:constraint-Phi_phi0_K=0_nondynamicalQ}\\
\end{equation}
while the dynamical equations reduce to
\begin{align}
 \partial_t \psi &= \alpha P + \beta\partial_r\psi\,, \label{eq:dpsidt_phi0_K=0_nondynamicalQ}\\
 \partial_t P &= \alpha \left[ \partial_r^2\psi + (a +2\partial_r\ln\Phi)\partial_r\psi\right] + \beta\partial_r P\,, \label{eq:dPdt_phi0_K=0_nondynamicalQ}\\
 \partial_t \Phi &= -\frac{1}{2}\alpha Q\Phi + \beta\partial_r\Phi\,, \label{eq:dPhidt_phi0_K=0_nondynamicalQ}
\end{align}
and the non-dynamical equations become
\begin{align}
 \partial_r \ln\alpha &= a\,, \label{eq:lapse_phi0_K=0_nondynamicalQ} \\
 \partial_r \beta &= - \alpha Q \,, \label{eq:shift_phi0_K=0_nondynamicalQ}\\
 \partial_r a &= - a^2  - 2a\partial_r\ln\Phi  +\frac{3}{2}Q^2 + P^2 \,,  \label{eq:eq_a_phi0_K=0_nondynamicalQ}\\
 \partial_r Q &= -3Q\partial_r\ln\Phi + P\partial_r\psi\,,\label{eq:Q_phi0_K=0_nondynamicalQ}
\end{align}
where equation \eqref{eq:Q_phi0_K=0_nondynamicalQ} comes from equation \eqref{eq:constraint-Q}. For convenience, equation \eqref{eq:constraint-Phi_phi0_K=0_nondynamicalQ} can be reformulated to define the quantity $\mathcal{C}$ as
\begin{equation}
 \mathcal{C} \equiv
 \partial_r^2\Phi + \frac{(\partial_r\Phi)^2-1}{2\Phi} + \left[ \frac{3}{8}Q^2 + \frac{1}{4}P^2 + \frac{1}{4}(\partial_r\psi)^2\right]\Phi\,.
 \label{eq:Definition_of_calC}
\end{equation}
In this case, if equation \eqref{eq:constraint-Phi_phi0_K=0_nondynamicalQ} is indeed well respected, the quantity $\mathcal{C}$ remains identically null.
Any deviations from $0$, which is bound to happen once numerically implemented, will allow us to watch and quantify the convergence of the numerical integration (see \cref{sec:numerical_integration}).

\subsection{Boundary conditions}
\label{sec:Boundary_conditions}

We need to be careful about the boundary condition at the center of spherical symmetry, where $\Phi=0$. By redefinition of the radial coordinate $r$, we set
\begin{equation}
    \Phi(t,r=0)=0 \,,\quad \partial_r\Phi(t,r=0)>0
\end{equation}
so that the center of spherical symmetry is at $r=0$ and that we consider the region $r\geq 0$. For the regularity of the center, we require ($\psi, P, Q, \alpha$) to be even functions of $r$ and $(\Phi, \beta, a)$ to be odd functions of $r$. Moreover, equations \eqref{eq:constraint-Phi_phi0_K=0_nondynamicalQ} and \eqref{eq:Q_phi0_K=0_nondynamicalQ} respectively imply that
\begin{equation}
    \partial_r\Phi(t,r=0)=1 \,,\qquad Q(t,r=0)=0 \, .
\end{equation}

By Taylor expanding the variables $(\psi, P, \Phi, Q, \alpha, \beta, a)$ with respect to $r$ around $r=0$ and inserting these, one can confirm that the right-hand sides of all relevant equations \eqref{eq:constraint-Phi_phi0_K=0_nondynamicalQ}-\eqref{eq:Q_phi0_K=0_nondynamicalQ} are well-defined in the limit $r\to +0$, implying that they can also be Taylor expanded with respect to $r$. As a consistency check, one can show that the right-hand sides of equations \eqref{eq:constraint-Phi_phi0_K=0_nondynamicalQ}, \eqref{eq:dpsidt_phi0_K=0_nondynamicalQ}, \eqref{eq:dPdt_phi0_K=0_nondynamicalQ}, \eqref{eq:shift_phi0_K=0_nondynamicalQ}, and \eqref{eq:eq_a_phi0_K=0_nondynamicalQ} are even functions of $r$, while the right-hand sides of equations \eqref{eq:dPhidt_phi0_K=0_nondynamicalQ}, \eqref{eq:lapse_phi0_K=0_nondynamicalQ}, and \eqref{eq:Q_phi0_K=0_nondynamicalQ} are odd functions of $r$. Furthermore, by using the $r\to +0$ limits of equations \eqref{eq:shift_phi0_K=0_nondynamicalQ} and \eqref{eq:constraint-Phi_phi0_K=0_nondynamicalQ} respectively, one can also show that the first $r$-derivative of the right-hand side of equation \eqref{eq:dPhidt_phi0_K=0_nondynamicalQ} vanishes at $r=0$.

We now introduce an outer boundary at $r=r_{\rm b}$ and hereafter consider the region $0\leq r \leq r_{\rm b}$. In order to evolve $(\psi, P, \Phi)$ with the dynamical equations \eqref{eq:dpsidt_phi0_K=0_nondynamicalQ}-\eqref{eq:dPhidt_phi0_K=0_nondynamicalQ}, we need to impose appropriate boundary conditions on the outer boundary as well. For example, if $r_{\rm b}$ is large enough then we can set
\begin{equation}
    \psi(t,r_{\rm b})=0 \,,\quad P(t,r_{\rm b})=0 \,,\quad \partial_r^2\Phi(t,r_{\rm b})=0 \,.
\end{equation}
Up to here, the overall normalization of $\alpha$ is not fixed, corresponding to the fact that the theory enjoys time reparametrization symmetry. We fix it by demanding that
\begin{equation}
    \alpha(t,r=r_{\rm b})=1
\end{equation}
so that the time $t$ agrees with the proper time measured by an observer at the outer boundary. The equations \eqref{eq:shift_phi0_K=0_nondynamicalQ} and \eqref{eq:eq_a_phi0_K=0_nondynamicalQ} for $\beta$ and the acceleration $a$ do not require additional boundary conditions; the fact that $\beta$ and $a$ are odd functions of $r$ already imposes $\beta(t,r=0)=0=a(t,r=0)$.

\subsection{Initial conditions}
\label{sec:Initial_condition}

We consider the following initial conditions at the initial time $t=t_0$:
\begin{equation}
 P(t_0,r) = 0\,,\quad Q(t_0,r) = 0\,,\quad \alpha(t_0,r) = 1\,, \quad \beta(t_0,r) = 0\,, \quad a(t_0,r) = 0\,,
\end{equation}
and
\begin{equation}
 \psi(t_0,r) = \psi_0(r)\,,\quad \Phi(t_0,r) = \Phi_0(r)\,,
\end{equation}
where $\psi_0(r)$ is a free function of $r$, which we, for concreteness, take to be an even Gaussian wave packet of the form
\begin{equation}
 \psi_0(r) = A \exp\left[ - \frac{(r^2-r_0^2)^2}{s^4}\right]\,.
 \label{eq:Initial_psi}
\end{equation}
The constants $A$, $r_0$ and $s$ determine the amplitude, starting position, and width of the initial wave packet respectively. Having defined $\psi_0$, the constraint \eqref{eq:constraint-Phi_phi0_K=0_nondynamicalQ} implies that $\Phi_0(r)$ is the solution to
\begin{equation}
 \frac{\partial_r^2\Phi_0}{\Phi_0} = \frac{1-(\partial_r\Phi_0)^2}{2\Phi_0^2} - \frac{1}{4}(\partial_r\psi_0)^2
\end{equation}
with the boundary conditions given in \cref{sec:Boundary_conditions}, \textit{i.e.}
\begin{equation}
 \Phi_0(r=0) = 0\,, \quad \partial_r\Phi_0(r=0) = 1\,.
\end{equation}

\section{Numerical integration}\label{sec:numerical_integration}

This section details the numerical approach to the problem. \Cref{sec:Numerical_setup} begins by explaining the schematics of the integration, explicitly stating the parameter choice for the initial conditions in \cref{sec:Initial_condition}, as well as different numerical remedies for taming arising instabilities. In \cref{sec:resolution} we present the simulation results and confirm, in \cref{sec:Apparent_horizon_formation}, the appearance of an apparent horizon followed by the evolution of the lapse function inside the horizon towards zero. Finally, we discuss the implications of the initial condition parameter choice on the dynamics and end results of the collapse in \cref{sec:parameter_variation}.

\subsection{Numerical setup}
\label{sec:Numerical_setup}

The dynamics of the collapse are fully described by three dynamical equations for $(\psi, P, \Phi)$ and four non-dynamical ones for $(\alpha, \beta, a, Q)$\footnote{These form a strongly hyperbolic system of partial differential equations, since after having integrated out the shadowy mode, they were shown to be the same as in \ac{gr}.}. We solve the system described in \cref{sec:Nondynamical_Q} with the boundary conditions and initial conditions formulated in \cref{sec:Boundary_conditions,sec:Initial_condition}. In order to remain consistent, the non-dynamical equations must be solved at every time evaluation. Starting from the initial time $t=t_0$, the initial condition must satisfy the constraint \eqref{eq:constraint-Phi_phi0_K=0_nondynamicalQ} and the non-dynamical equations \eqref{eq:lapse_phi0_K=0_nondynamicalQ}-\eqref{eq:Q_phi0_K=0_nondynamicalQ}. We then evolve $(\psi, P, \Phi)$ to the next time step $t=t_1$ with the dynamical equations \eqref{eq:dpsidt_phi0_K=0_nondynamicalQ}-\eqref{eq:dPhidt_phi0_K=0_nondynamicalQ}. In order to obtain $(\alpha, \beta, a, Q)$ at $t=t_1$, we integrate the non-dynamical equations \eqref{eq:lapse_phi0_K=0_nondynamicalQ}-\eqref{eq:Q_phi0_K=0_nondynamicalQ} once more, now on the time slice $t=t_1$. At this point, the numerical accuracy of the solution may be gauged by checking how well the data on the time slice $t=t_1$ satisfies the constraint \eqref{eq:constraint-Phi_phi0_K=0_nondynamicalQ}. The numerical scheme can then be advanced to the next time step by repeating this procedure.

Unless stated otherwise, the initial field $\psi$ is given by an even Gaussian wave packet (defined in equation \eqref{eq:Initial_psi}) of height $A = 0.15$ and a width characterized by $s = 4$ positioned at $r_0 = 10$. An outer boundary $r_{\mathrm{b}}$ of the system is placed far from the origin $r=0$ and the initial wave packet, to prevent it from disturbing the collapsing process and to best simulate an asymptotically flat spacetime. In our case, $r_{\mathrm{b}} = 80$ was used. Larger or slightly smaller values yield equivalent results. Space and time are uniformly discretized into intervals of size $\Delta r$ and $\Delta t$, respectively. We decide here to keep the ratio $\Delta t / \Delta r = 0.2$ constant. A smaller ratio would also have been acceptable, but if $\Delta t$ is too large relative to $\Delta r$, the system may become unstable.

To prevent any numerical instabilities from jeopardizing the computations, we also adopt two additional safety measures. Firstly\,---\,and most importantly\,---\,we use a series expansion near the boundaries to approximate the first and last integration points of the nondynamical fields (see \cref{appendix:B}). Secondly, we include Kreiss-Oliger dissipation terms \cite{KreissOligerBook}. These two measures serve to mitigate any possible emergent high-frequency modes near the boundaries and around the scalar field profile. As is standard, the artificial dissipation is chosen to be two orders higher than the convergence order of the spatial numerical derivatives.

\begin{figure}[H]
    \centering
    \includegraphics{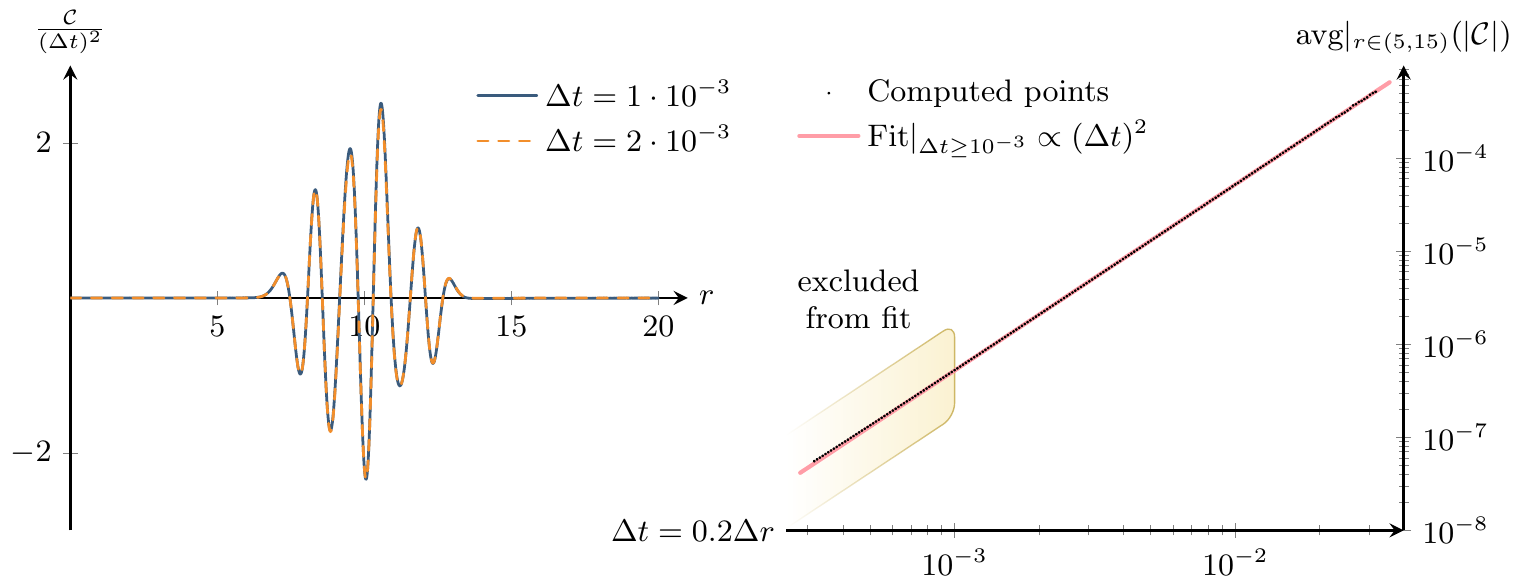}
    \caption{
        The left plot displays the constraint $\mathcal{C}$, as defined in equation \eqref{eq:Definition_of_calC}, divided by the square of the time step $\Delta t$, versus the radius $r$, for two different values of $\Delta t$. We reiterate that $\Delta t=0.2\Delta r$. As the shape remains unchanged for the two values of $\Delta t$, we conclude that $\mathcal{C} \propto (\Delta t)^2$ and that the numerical convergence is of quadratic order as expected.
        The right plot portrays the average absolute value of $\mathcal{C}$ for $5\leq r\leq 15$ (the region in the left plot for which $\mathcal{C}$ clearly is non-zero) against $\Delta t$. A simple linear fit neatly illustrates the quadratic convergence order.
        The fit is here restricted to $\Delta t \geq 10^{-3}$ values in order to avoid the small numerical noise that starts seeping in at lower step sizes (the excluded points are highlighted above).
        Both plots are generated at time $t = 2$. Albeit early after the numerical resolution is engaged, choosing different times $t$ does not affect the convergence order.
        \label{fig:Convergence_plots}
    }
\end{figure}

The simulation was executed with several independently written codes, using different integration schemes, and all results matched.
As an extra sanity check, the quantity $\mathcal{C}$, defined in equation \eqref{eq:Definition_of_calC}, was continuously and carefully surveilled to ensure it remained sufficiently small throughout the simulation. A more quantitative examination of $\mathcal{C}$ was also carried out, as \cref{fig:Convergence_plots} demonstrates. While the shape of $\mathcal{C}$ remains unchanged (left plot), its overall amplitude is inversely proportional to the square of the time step $\Delta t$ or, equivalently, the spatial gap $\Delta r$ (right plot).
The quadratic convergence is expected. All codes\,---\,two of which are, in fact, the same codes as in \cite{Berti:2022xfj, Martens:2022dtd} and one of the two was also used in \cite{Bhattacharjee:2018nus,Mukohyama:2020lsu}\,---\,exploit second-order methods to integrate the non-dynamical equations in space, and at least second-order methods for the integration in time. Explicitly, one used the trapezoidal method in space and the iterated Crank-Nicolson method with two iterations in time, while the other two used Heun's method in space and higher order methods, mainly RK4, in time.
\subsection{Results}\label{sec:resolution}

\begin{figure}[H]
    \centering
    \includegraphics{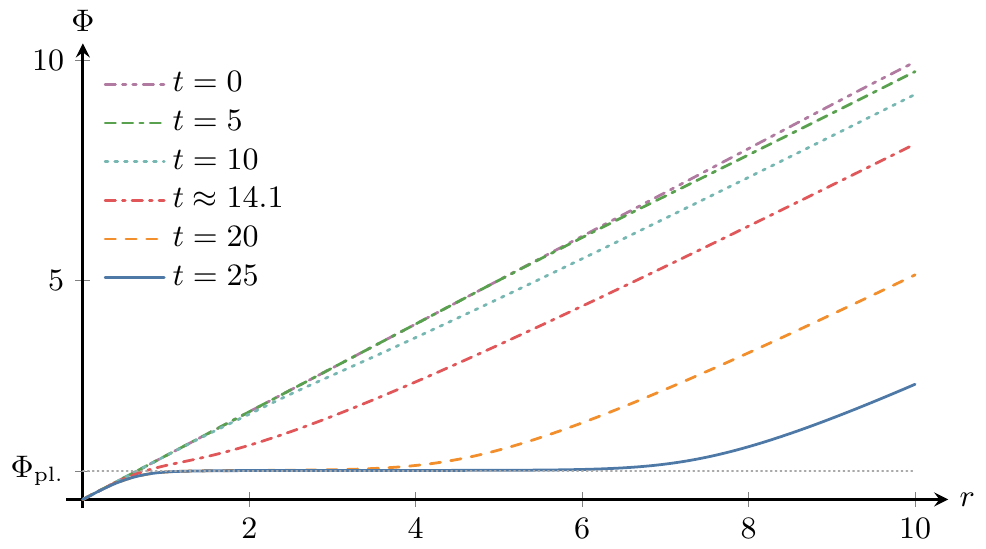}
    \caption{The areal radius $\Phi$ plotted against the proper distance $r$ from the center at different times $t$. One may observe the boundary conditions established in \cref{sec:Boundary_conditions} in the shape of $\Phi$. Following the fall of the right side of $\Phi$, a plateau forms around $\Phi_{\text{pl.}} \approx 0.65$.
    \label{fig:Phi_wrt_r}}
\end{figure}

We now turn to the physical quantities thusly obtained\footnote{We remind the reader that as a consequence of having integrated out the shadowy mode, the integrated equations \eqref{eq:constraint-Phi_phi0_K=0_nondynamicalQ}-\eqref{eq:Q_phi0_K=0_nondynamicalQ} are the same as in \ac{gr}. Naturally, the results are then the same as in \ac{gr} as far as the foliation-independent parts are concerned. Unlike in \ac{gr}, however, the foliation-dependent parts are also physical in VCDM and thus will be investigated with care.}. \Cref{fig:Phi_wrt_r} shows the areal radius $\Phi$ against the proper distance $r$ for some selected times $t$.
Notice that one of the displayed times, in this figure and the subsequent ones, is $t \approx 14.1$. As we shall explain in \cref{sec:Apparent_horizon_formation}, this corresponds to the \ac{ah} formation time.
At $t=0$, we observe that $\Phi \approx r$. As time proceeds, it only retains this behavior near $r=0$ (in accordance with \cref{sec:Boundary_conditions}), falling and settling down into a plateau elsewhere.
The height of this plateau, $\Phi_{\text{pl.}} \approx 0.65$, echoes in the other quantities, as the subsequent figures show.
Henceforth, we shall use $\Phi$ as a radial coordinate to illustrate the evolution of all other quantities.

\begin{figure}[H]
    \centering
    \includegraphics{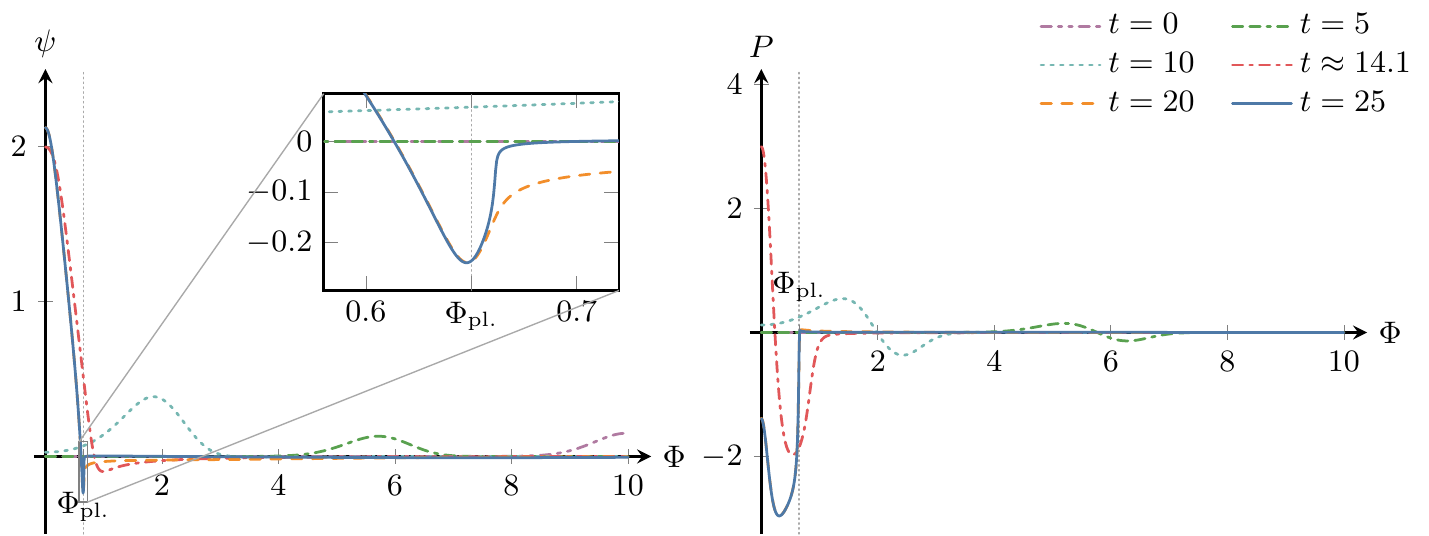}
    \caption{The scalar field $\psi$ (left) and $P$ (right) against the areal radius $\Phi$ for different values of time $t$. The two sharp changes seen at $t=25$\,---\,the dip under zero for $\psi$ and the sharp turn for $P$\,---\,are located around $\Phi=\Phi_{\text{pl.}}$. The former is further detailed in the inset plot, where one can observe that the apparent spike is smooth. For the chosen range of $\Phi$, the split of $\psi$ is not explicitly observable (as $r_0 = 10$).
    \label{fig:psi_and_P_wrt_Phi}}
\end{figure}

The other two dynamical quantities $\psi$ and $P$ are given for different times in \cref{fig:psi_and_P_wrt_Phi}. For the prescribed initial conditions, the scalar field $\psi$ starts by splitting into two parts. One moves out to $r=\infty$ and vanishes, while the other falls to the origin $r=0$ under its self-gravity and eventually settles near $r=0$. Naturally, this foreshadows an \ac{ah} formation (see \cref{sec:Apparent_horizon_formation}). A similar behavior is observed for the related variable $P$ ($\equiv \partial_\perp \psi$, as defined in equation \eqref{eq:Definition_of_Q_P_and_a}).

\begin{figure}[H]
    \centering
    \includegraphics{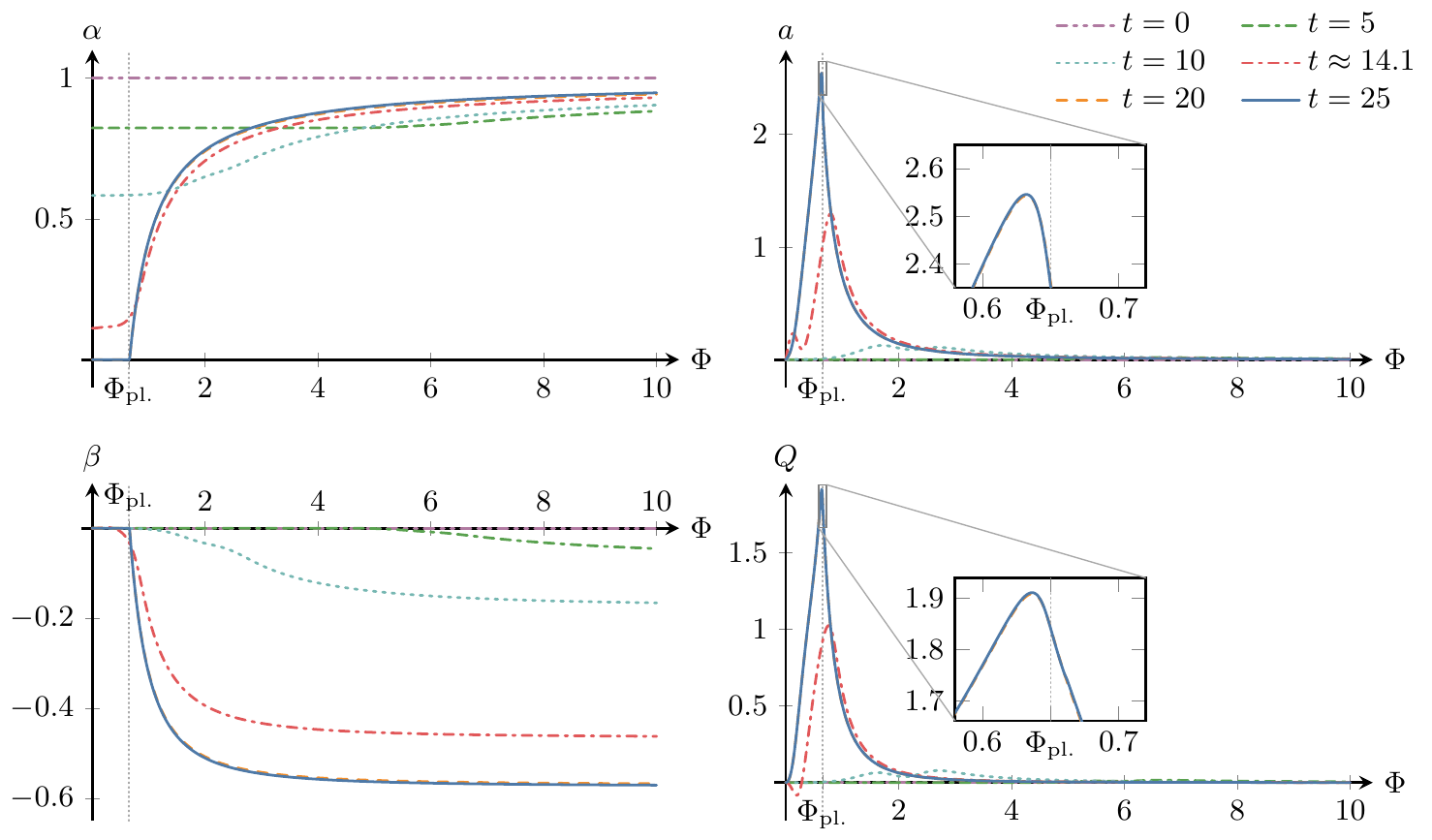}
    \caption{All four nondynamical fields $\alpha$, $a$, $\beta$ and $Q$, here plotted against the areal radius $\Phi$ at different times $t$. Similarly to \cref{fig:psi_and_P_wrt_Phi}, two inset plots for $a$ and $Q$ were added to explicitly show the peak smoothness. \label{fig:Nondynamical_fields}}
\end{figure}

In parallel, the results for the nondynamical fields are given in \cref{fig:Nondynamical_fields}.
One can observe how the lapse $\alpha$ and the shift $\beta$ collapse to zero below the threshold value of $\Phi_{\text{pl.}}$. Far from the origin, an asymptotically flat value is manifestly recovered for both quantities.
The variables $a$ and $Q$ simultaneously spike around the same threshold value of $\Phi_{\text{pl.}}$.

\subsection{Apparent horizon formation}
\label{sec:Apparent_horizon_formation}

\begin{figure}[H]
    \centering
    \includegraphics{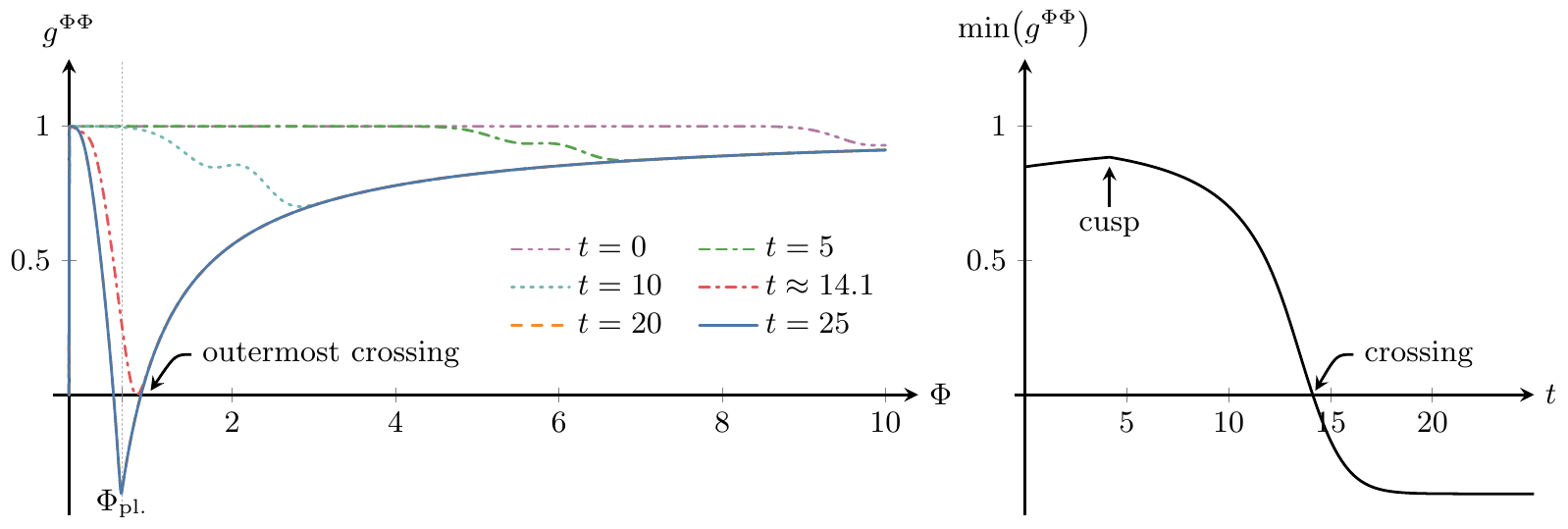}
    \caption{The left panel gives $g^{\Phi\Phi}$ with respect to the areal radius $\Phi$ at different times and demonstrates the creation of an \ac{ah} (see equation \eqref{eq:AH_condition}). At late time, $g^{\Phi\Phi}$ settles with a minimum around $\Phi_{\text{pl.}}$ and an outermost crossing near $\Phi \approx 0.88$. The right panel details the minimum of $g^{\Phi\Phi}$ as a function of time and portrays how the horizon forms at $t \approx 14.1$, at the \enquote{crossing}. A careful reader may wonder about the small cusp near $t\approx 4.1$: it is a consequence of $\psi$ splitting into two at the beginning of the simulation.
    \label{fig:gPhiPhi_wrt_Phi}}
\end{figure}

The quantity $g^{\Phi\Phi}$ (defined in equation \eqref{eq:gPhiPhi_without_K}) is depicted for different times in the left plot of \cref{fig:gPhiPhi_wrt_Phi}.
As the behavior of $\psi$ already suggested (see \cref{fig:psi_and_P_wrt_Phi}), the collapse does in fact yield an \ac{ah}, which occurs when $g^{\Phi\Phi}$ crosses zero.
This condition is satisfied at $t \approx 14.1$, as the right plot of \cref{fig:gPhiPhi_wrt_Phi} shows. Furthermore, in \cref{fig:Nondynamical_fields}, one can deduce that $\alpha$ and $\beta$ have not yet collapsed down to $0$ at the time of the crossing, implying that the \ac{ah} indeed forms before any singularity or breakdown of the time slicing does. Eventually, as the lapse and shift evolve towards zero inside the \ac{ah}, $g^{\Phi\Phi}$ settles with a minimum around $\Phi_{\text{pl.}} \approx 0.65$ and an outermost crossing of zero at $\Phi \approx 0.88$. This constitutes the main achievement of this work, namely that the gravitational collapse of a massless scalar field in VCDM results in the manifestation of an \ac{ah} before the breakdown of the time foliation. The final stage of the collapse is a static black hole.

\subsection{Parameter variation}\label{sec:parameter_variation}

\begin{figure}[H]
    \centering
    \includegraphics{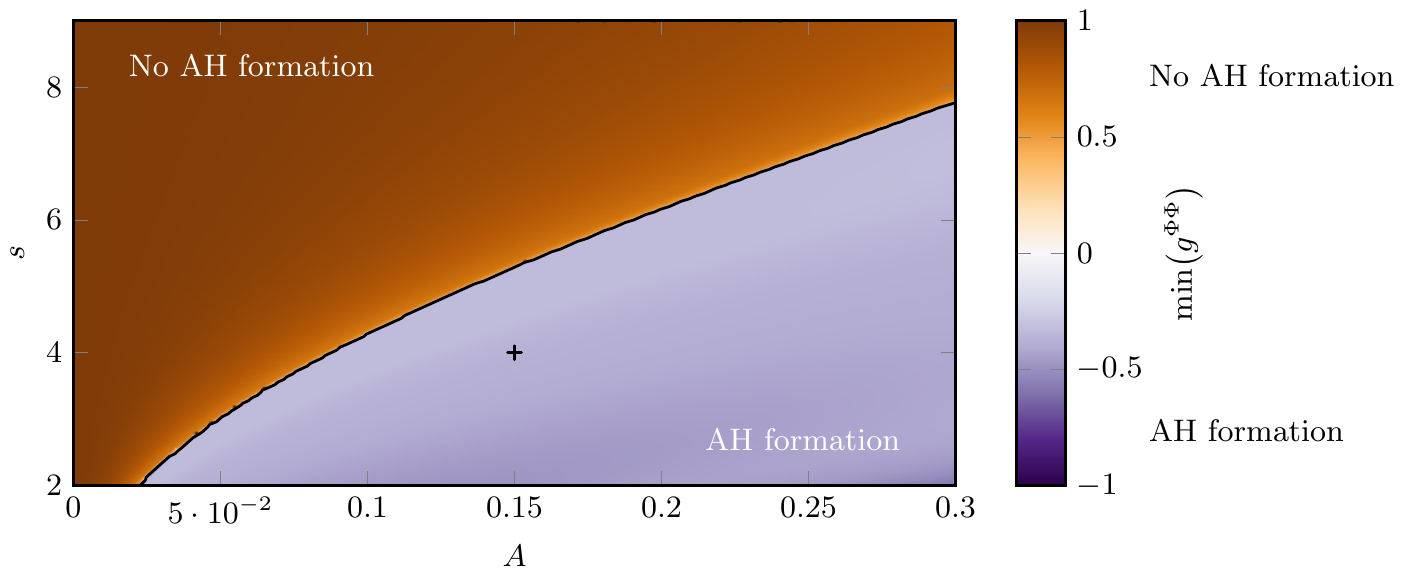}
    \caption{
        The minimal value (in time and space) that $g^{\Phi\Phi}$ reaches before $t=30$, which, given that the initial wave packet of $\psi$ is placed at $r_0=10$, is enough time to determine whether a horizon forms or not (see equation \eqref{eq:AH_condition}).
        The black cross indicates the $A=0.15$ and $s=4$ parameter pair that was previously used, while the black line approximately delimits $\min(g^{\Phi\Phi}) = 0$.
        The range in $A$ was uniformly sampled in $301$ points from $0.0$ to $0.3$, while the values in $s$ were sampled in $176$ points uniformly spread from $2.0$ to $9.0$.
        \label{fig:mingPhiPhi_wrt_A_and_s}}
\end{figure}

\begin{figure}[H]
    \centering
    \includegraphics{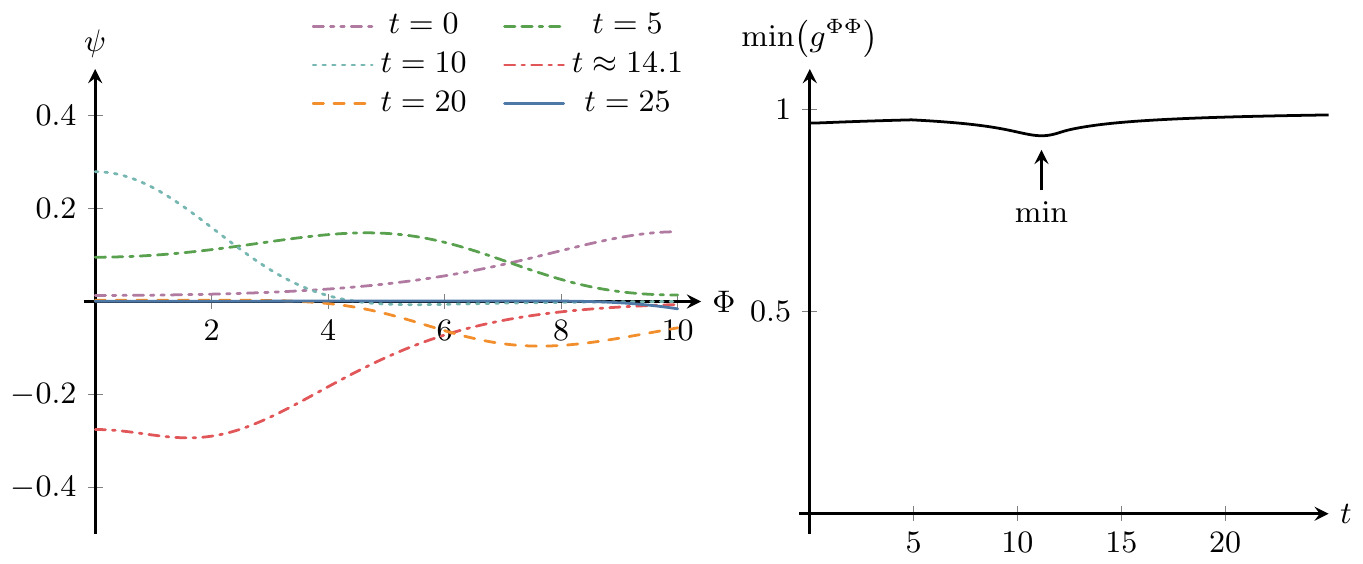}
    \caption{
        Evolution of the scalar field $\psi$ for $A=0.15$ and $s=8$ on the left.
        Observe how the packet moves toward the origin before bouncing back.
        At $t=25$, the bulk is mostly out of the represented range.
        The right plot gives $\min(g^{\Phi\Phi})$ for the same configuration. Naturally, no \ac{ah} forms (\textit{i.e}. $\forall t:\,g^{\Phi\Phi} > 0$).
        \label{fig:No_horizon_illustration}
    }
\end{figure}

Until now, we have kept fixed the initial scalar field parameters of $\psi_0$ as defined in equation \eqref{eq:Initial_psi} to $A=0.15$ and $s=4$.
However, depending on the values assigned to these two parameters, an \ac{ah} may or may not form.
The parameter space where an \ac{ah} does appear is illustrated in \cref{fig:mingPhiPhi_wrt_A_and_s}.
A decrease in $A$ and/or an increase of $s$ may both \enquote{dilute} the scalar field enough to allow it to bounce back around the origin before any \ac{ah} appears or the time slicing breaks down.
An example of this is shown in the left plot of \cref{fig:No_horizon_illustration}, where we have chosen $s=8$ and plotted $\psi$ at different times throughout the evolution. The right plot of the same figure makes the non-formation of an \ac{ah} evident: $g^{\Phi\Phi}$ never crosses zero.

\section{Summary and discussion} \label{sec:summary_and_discussion}

As is well-known, gravitational collapse in the framework of \ac{gr} leads to the formation of a black hole, a trapped region of spacetime bounded by a marginally outer trapped surface, called the apparent horizon. Deviations from this picture may lead to contradictions through various observations such as \ac{gws} and black hole shadows, and so whether an alternative theory of gravity admits black holes or not is often used as a litmus test for its validity. It has been previously shown in \cite{DeFelice:2022uxv} that a subset of the solutions of VCDM coincides exactly with solutions in \ac{gr} when the 3+1-decomposition admits a constant trace of the extrinsic curvature. This subset of VCDM solutions consists of those for which the auxiliary field $\phi$ and Lagrange multiplier $\lambda$ are constant. Reassuringly, \cite{DeFelice:2022riv} shows that the spherical gravitational collapse of a homogeneous cloud of dust in VCDM coincides with a foliation of the Oppenheimer-Snyder collapse in GR for which indeed the aforementioned criteria are satisfied.

In the continuation of \cite{DeFelice:2022riv}, we have numerically investigated the collapse of a massless scalar field in VCDM, in hopes of observing the formation of a black hole and whether its horizon appears before or after any singularity or foliation breakdown. Starting from a spherically symmetric ansatz of the metric and other relevant quantities in the total action \eqref{eq:total_action}, we derived the equations of motion. By requiring vacuum flat spacetime in the standard Minkowski time slicing at infinity and regularity at $r=0$, we have shown $\phi$ to be a constant and the trace of the extrinsic curvature to vanish. An immediate consequence, as implied by equation \eqref{eq:EoM_lambdas}, was that $\lambda$ also reduced to a constant. Hence, the sufficient conditions were satisfied for solutions of this system to coincide with ones in \ac{gr} in the specified time foliation.

The equations of motion reduce to a constraint equation \eqref{eq:constraint-Phi_phi0_K=0_nondynamicalQ}, three dynamical equations \eqref{eq:dpsidt_phi0_K=0_nondynamicalQ}-\eqref{eq:dPhidt_phi0_K=0_nondynamicalQ}, and four non-dynamical ones \eqref{eq:lapse_phi0_K=0_nondynamicalQ}-\eqref{eq:Q_phi0_K=0_nondynamicalQ}. These were integrated with the boundary conditions in \cref{sec:Boundary_conditions} and initial conditions in \cref{sec:Initial_condition}. Furthermore, the quantity $\mathcal{C}$ (equation \eqref{eq:Definition_of_calC}) was used for assessing the numerical accuracy of the integration. The formation of an \ac{ah} was shown through $g^{\Phi\Phi}$ as defined in equation \eqref{eq:gPhiPhi_without_K} and by checking the condition \eqref{eq:AH_condition}. To ensure numerical stability, the meshing in time and space was chosen such that $\Delta t/\Delta r=0.2$. Moreover, emergent high-frequency oscillations were tamed by implementing Taylor expansions near the boundaries and Kreiss-Oliger dissipation terms.

First and foremost, as all three independent codes utilized second-order methods in space, it is reassuring that \cref{fig:Convergence_plots} confirms a quadratic convergence of the error. Secondly, as seen in \cref{fig:gPhiPhi_wrt_Phi}, the solutions to equations \eqref{eq:dpsidt_phi0_K=0_nondynamicalQ}-\eqref{eq:Q_phi0_K=0_nondynamicalQ} indeed lead to the formation of an \ac{ah}. Concurrently considering \cref{fig:Nondynamical_fields}, the lapse $\alpha$ and shift $\beta$ are everywhere non-zero at the time of the \ac{ah} formation, meaning the time foliation preferred by the theory fully describes spacetime inside the black hole at the time of its formation. As the simulation proceeds, $\alpha$ and $\beta$ converge towards zero in a finite region inside\,---\,but not up to\,---\,the \ac{ah}. The proper time in this region elapses slower and slower than that measured by an observer far away at infinity as the universe evolves, eventually standing still. In other words, a breakdown of the time foliation inside the \ac{ah} is in the process, though from the viewpoint of observers outside the horizon, it will take infinitely long to actually happen. This is by all means not an issue, as there is also an infinite amount of time for the universe to evolve in VCDM. It does however imply that a UV-completion of the theory is needed to fully describe the inside of the black hole.

Further testing of VCDM is necessary to probe its full potential. As previously mentioned, \cite{DeFelice:2022uxv} tested the corresponding Oppenheimer-Snyder case; this paper proceeded with studying a collapsing massless scalar field. We have observed the creation of a black hole, its \ac{ah} forming prior to the breakdown of the time foliation and the formation of the singularity. In the future, one could consider a collapsing star made up of more realistic matter, such as a fluid, or adding further layers of properties, \textit{e.g.} charge or angular momentum. While this would drastically complicate the dynamics of the collapse, it would certainly provide useful insight into the viability of VCDM.

Any form of non-symmetrical dynamics\,---\,rotating black holes, black hole and neutron star binaries, \ac{gws}\,---\,may prevent $\phi$ from being constant in time. This would yield non-\ac{gr} solutions of VCDM, which necessarily need to be studied if the theory is to be fully explored.

\acknowledgments
P.M. acknowledges support from the Japanese Government (MEXT) scholarship for Research Student. A.F.J. acknowledges support from the Sweden Japan Foundation as well as from JASSO for international exchange students coming to Japan. The work of S.M. was supported in part by World Premier International Research Center Initiative, MEXT, Japan.

\appendix
\section{Apparent horizon}\label{appendix:A}

In this appendix we derive the equations \eqref{eq:AH_condition} and \eqref{eq:Definition_of_gPhiPhi}, starting from a spherically symmetric ansatz in the ADM formalism as the one in \cref{sec:basic_equations},
\begin{align}
    g_{\mu\nu}dx^{\mu}dx^{\nu}=-\alpha^2dt^2 + (dr+\beta dt)^2 + \Phi^2d\Omega^2\,,
\end{align}
where $\alpha,\beta$ are positive functions of $t$ and $r$, and $\Phi=\Phi(t,r)$ is non-negative. Equation \eqref{eq:Definition_of_gPhiPhi} can be found immediately by a redefinition of the radial coordinate $r\rightarrow \Phi(t,r)$:
\begin{align}
\begin{split}
    g^{\Phi\Phi}&=\dfrac{\partial \Phi}{\partial x^{\mu}}\dfrac{\partial \Phi}{\partial x^{\nu}}g^{\mu\nu}\\
    &= -(\partial_{\perp}\Phi)^2+(\partial_r\Phi)^2\,,\label{eq:gPhiPhi_appendix}
\end{split}
\end{align}
where $\partial_{\perp}=(\partial_t-\beta\partial_r)/\alpha$ is the derivative in the direction perpendicular to the surfaces of the slicing. Although $\Phi$ will not be used as a radial coordinate in what follows, the above derivation will prove convenient shortly. Let us construct a pair of null vector fields $k$ and $l$:
\begin{align}
    k^{\mu}\partial_{\mu}\equiv \partial_{\perp} + \partial_r\,,\\
    l^{\mu}\partial_{\mu}\equiv \partial_{\perp} - \partial_r\,.
\end{align}
The vector field $k$ is tangent to a congruence of radially outgoing null curves, and similarly, $l$ is tangent to a congruence of radially ingoing ones. We acknowledge the relations
\begin{align}
k^{\mu}k_{\mu}=l^{\mu}l_{\mu}=0,\quad k^{\mu}l_{\mu}=l^{\mu}k_{\mu}=-2\,.
\end{align}
The part of $g^{\mu\nu}$ transverse to these two vector fields, which we will denote by $\sigma^{\mu\nu}$, is given by
\begin{align}
\sigma^{\mu\nu}=g^{\mu\nu}+\dfrac{1}{2}(k^{\mu}l^{\nu}+l^{\mu}k^{\nu})\,.
\end{align}
This metric has components $\sigma^{tt}=\sigma^{tr}=\sigma^{rt}=\sigma^{rr}=0$. The apparent horizon $r=r_{\text{AH}}$ is found by solving for a marginally outer trapped surface of the spacetime, that is, by finding when the expansion scalar $\theta_{\text{out}}$ of the outgoing congruence, i.e. with tangent field $k$, is $0$. The expansion scalar $\theta_{\text{out}}$ is given by the transverse projection of the covariant derivative of $k$:
\begin{align}
    \theta_{\text{out}}\equiv \sigma^{\mu\nu}\nabla_{\nu}k_{\mu}=\dfrac{2}{\Phi}(\partial_{\perp}\Phi+\partial_r\Phi)\,.
\end{align}
If we now set $\theta_{\text{out}}\vert_{r=r_{\text{AH}}}=0$, it follows that
\begin{align}
 (\partial_{\perp}\Phi+\partial_r\Phi)\big\vert_{r=r_{\text{AH}}} = 0\,.
\end{align}
This implies the equality
\begin{align}\label{eq:AH_condition_appendix}
    g^{\Phi\Phi}\big\vert_{r=r_{\text{AH}}}=0\,,
\end{align}
where $g^{\Phi\Phi}$ is given in equation \eqref{eq:gPhiPhi_appendix}. Actually, the following identity can be easily derived:
\begin{align}
    g^{\Phi\Phi} = -\dfrac{\Phi^2}{4} \theta_{\text{in}}\theta_{\text{out}}\,,
\end{align}
where
\begin{align}
    \theta_{\text{in}}\equiv \sigma^{\mu\nu}\nabla_{\nu}l_{\mu}=\dfrac{2}{\Phi}\left(\partial_{\perp}\Phi-\partial_r\Phi\right)\,,
\end{align}
and one can numerically confirm that $\theta_{\text{in}}$ does not vanish during the collapse. Consequently, if an apparent horizon exists, its position is given by a solution to equation \eqref{eq:AH_condition_appendix}. One can easily show that $\theta_{\text{in}}<0<\theta_{\text{out}}$ at the regular origin $\Phi=0$ and in the asymptotically flat region $\Phi\to\infty$. Therefore, the roots of $g^{\Phi\Phi}$ should be pairwise as long as they are not degenerate. The outer most root of $g^{\Phi\Phi}$ corresponds to the apparent horizon. It is then simply a matter of substitution using \eqref{eq:Definition_of_Q_P_and_a} to arrive at the right-hand side expression seen in equation \eqref{eq:Definition_of_gPhiPhi}.

\section{Taylor expansions near the boundaries}\label{appendix:B}

We will now discuss the Taylor expansions that were implemented near the boundaries to prevent high-frequency oscillations from appearing. This was only applied to the non-dynamical fields, as the dynamical fields were known in all of space at every time step.

Suppose we have $N$ dynamical fields $f_n(x)$ and $M$ non-dynamical fields $g_m(x)$, where $n=1,\ldots,N$ and $m=1,\ldots,M$, and $x$ is the spatial variable discretized with step-length $\Delta x$. Any quantity $h(x)$ can be Taylor-expanded to some order $k_{\rm max}$ around some point $x_0$ accordingly,
    \begin{align}
        h(x)=\sum_{k=0}^{k_{\rm max}}\dfrac{h^{(k)}(x_0)}{k!}(x-x_0)^k+\mathcal{O}(x^{k_{\rm max}+1})\,.
    \end{align}
The following procedure is applicable to any point $x_0$.
\begin{enumerate}
    \item
    Begin by Taylor expanding all quantities, both dynamical and non-dynamical, to a sufficiently high order $k_{\rm max}$ around $x_0$. The order generally depends on how many points one wishes to manually set. If $x_0$ is a boundary, then apply relevant boundary conditions by adjusting the coefficients adequately.

    \item Suppose we are considering $k_{\rm max}+1$ points $x_j$ in the closest vicinity of $x_0$ where $j=0, 1,\ldots,k_{\rm max}$ (which includes $x_0$). Since the dynamical fields $f_n$ are known at each point, their expansion coefficients $f_n^{(k)}(x_0)$ are easily computed by equating the given value of a dynamical field $f_n$ at $x_j$ with its respective Taylor expansion evaluated at the same point. This will yield a system of $k_{\rm max}+1$ equations,
    \begin{align}
        \left\{f_n(x_j) \approx \sum_{k=0}^{k_{\rm max}}\dfrac{f_n^{(k)}(x_0)}{k!}(x_j-x_0)^k : j = 0, 1, \dots, k_{\rm max}+1 \right\}\,,
    \end{align}
    that can be solved for the $k_{\rm max}+1$ unknown expansion coefficients $f_n^{(k)}(x_0)$ in terms of the known values $f_n(x_j)$. Let us denote their thus computed value as $\overline{f}_n^{(k)}$.
    \item Insert the now fully numerical expressions $\overline{f}_n^{(k)}$ for the $f_n^{(k)}(x_0)$ into the Taylor expansions of $f_n(x)$:
    \begin{align}
        f_n(x) = \sum_{k=0}^{k_{\rm max}}\dfrac{\overline{f}_n^{(k)}}{k!}(x-x_0)^k+\mathcal{O}(x^{k_{\rm max}+1})\,.
    \end{align}

    \item Insert all Taylor expansions $f_n(x)$ and $g_m(x)$ into the non-dynamical equations and gather together terms of each order. The coefficients in front of the terms will provide equations relating the non-dynamical expansion coefficients $g_m^{(k)}(x_0)$ to the fully numerical $\overline{f}_n^{(k)}$. Solve these equations to acquire fully numerical solutions $\overline{g}_m^{(k)}$ for the non-dynamical expansion coefficients $g_m^{(k)}(x_0)$.

    \item Insert the fully numerical solutions $\overline{g}_m^{(k)}$ for the expansion coefficients $g_m^{(k)}(x_0)$ into the Taylor expansions of the non-dynamical fields $g_m(x)$. We now have fully determined polynomial expressions for the non-dynamical fields around $x_0$:
    \begin{align}
        g_m(x)\approx \sum_{k=0}^{k_{\rm max}}\dfrac{\overline{g}_m^{(k)}}{k!}(x-x_0)^{k}\,.
    \end{align}
    \item Finally, manually set the value of $g_m$ in any point close to $x_0$ by simply evaluating its expansion in the said point.
\end{enumerate}

In our case, this procedure was applied to the points $r=0$ and $r=r_b$, where we set the values of the dynamical fields in the first three points and the last two points of the integration space $[0,r_b]$.

\bibliography{bibliography.bib}

\end{document}